\documentclass[10 pt, conference]{ieeeconf}  

\IEEEoverridecommandlockouts                              
\usepackage{graphicx} 
\usepackage{amsmath} 
\usepackage{amssymb}  
\usepackage{mathrsfs}
\usepackage{dsfont} 
\usepackage{bbm}
\usepackage{xcolor}

\newtheorem{thm}{Theorem}[section]
\newtheorem{lem}[thm]{Lemma}
 
\newtheorem{defi}[thm]{Definition}
\newtheorem{nota}[thm]{Notation}

\title{\LARGE \bf
On Exponential Stabilization of Spin-$\frac{1}{2}$ Systems
}

\author{Weichao Liang, Nina H. Amini, and Paolo Mason
\thanks{All authors are with Laboratoire des Signaux et Syst\`emes, CNRS - CentraleSup\'elec - Univ. Paris-Sud, Universit\'e Paris-Saclay, 3, rue Joliot Curie, 91192, Gif-sur-Yvette, France. {\tt\small [first name].[family name]@l2s.centralesupelec.fr}}%
}

\begin{document}

\maketitle
\thispagestyle{empty}
\pagestyle{empty}

\begin{abstract}
In this paper, we study the stabilization problem of quantum spin-$\frac{1}{2}$ systems under continuous-time measurements. In the case without feedback, we show exponential stabilization around the excited and ground state by providing a lower bound of the convergence rate. Based on stochastic Lyapunov techniques, we propose a parametrized feedback controller ensuring exponential convergence toward the target state. Moreover, we provide a lower bound of the convergence rate for this case. Then, we discuss the effect of each parameter appearing in the controller in the convergence rate. Finally, we illustrate the efficiency of such feedback controller through simulations.
\end{abstract}
\section{Introduction}
The evolution of an open quantum system undergoing indirect continuous-time measurements is described by the so-called quantum stochastic master equation, which has  been derived by Belavkin in quantum filtering theory \cite{belavkin1989nondemolition} (see also \cite{bouten2007introduction, van2005feedback} for a modern treatment). The solutions of such equation are called quantum trajectories 
and their properties have been studied in \cite{mirrahimi2007stabilizing, pellegrini2008existence}.\\
The deterministic part of quantum stochastic master equation, which corresponds to the average dynamics, is given by the well known Lindblad operator. Its stochastic part represents the back-action effect of continuous-time measurements. Feedback control of open quantum systems has been the subject of many papers \cite{van2005feedback, mirrahimi2007stabilizing, tsumura2008global, ge2012non}. In all of these papers, the control input appears in the Lindblad operator (through the system Hamiltonian). The experiment considered in \cite{sayrin2011real} is a relevant example where a  real-time feedback control is applied to stabilize an arbitrary photon number state in a microwave cavity. \\ 
The preparation of a pure state is investigated as an essential step towards quantum technologies. According to  \cite{abe2008analysis}, the stochastic part, unlike the deterministic one, contributes to increase the purity of the quantum state, thus yielding a desirable effect for the preparation of a pure state. On the other hand, if we do not implement a control to the quantum system, the measurement induces a collapse of the quantum state towards either one of the eigenstates of the measurement operator, a phenomenon known as quantum state reduction \cite{adler2001martingale, van2005feedback, mirrahimi2007stabilizing, sarlette2017deterministic}. Thus, combining the continuous measurement with the feedback control may provide a desirable approach for preparing the target state in practice.\\
In \cite{van2005feedback}, the authors design for the first time a quantum feedback controller that globally stabilizes a quantum spin-$\frac{1}{2}$ system (which is a special case of quantum angular momentum systems) towards an eigenstate of $\sigma_z$ in the presence of imperfect measurements. This feedback controller has been designed by looking numerically for an appropriate global Lyapunov function. Then, in \cite{mirrahimi2007stabilizing}, by analyzing the stochastic flow and by using stochastic Lyapunov techniques, the authors construct a switching feedback controller which globally stabilizes  the $N$-level quantum angular momentum system, in the presence of  imperfect measurements, to the target eigenstate. A continuous version of this feedback controller has been proposed in \cite{tsumura2008global}. The essential ideas in \cite{van2005feedback, tsumura2008global} for constructing the continuous feedback controller remain the same: the controllers consist of two parts, the first one making the state converge locally to the target eigenstate, and the second one driving the system away from the antipodal eigenstates. More recently, \cite{ge2012non} introduces a non-smooth Lyapunov-like theory to construct a continuous feedback controller that stabilizes globally the generic $N$-level quantum system at the target eigenstate.\\ 
The problem of estimating and optimizing the rate of convergence to the target eigenstate is also relevant in order to make the preparation more robust. For example, in \cite{benoist2017exponential}, the authors prove  the exponential stability of particular target subspaces for general $N$-level quantum system driven by Wiener processes and Poisson processes with an open-loop control (time-depending Hamiltonian) and in the case of perfect measurements. Also, in the recent paper \cite{cardona2018exponential}, the authors have proven, by simple
Lyapunov arguments, the stochastic exponential  stability for a
related control system obtained by applying a proportional output
feedback.\\
The contribution of this paper is threefold: Firstly, we prove exponential stabilization around the excited state and ground state for a spin-$\frac{1}{2}$ system without feedback and we provide a lower bound of the convergence rate; Secondly, we construct a parametrized feedback controller which almost surely stabilizes exponentially the spin-$\frac{1}{2}$ system around the excited state (or ground state) providing a lower bound of the convergence rate;  Finally, we study the effects of the parameters of the controller on the rate of convergence.\\ 
This paper is organized as follows. In Section~\ref{sec-pre}, we introduce the stochastic model describing spin-$\frac12$ systems containing control inputs in the presence of imperfect measurements and we introduce the notions of stochastic stability needed throughout the paper. In Section~\ref{sec-qsr}, we analyze the system without feedback and we provide an estimation of the rate of convergence to the excited or ground state in the quantum state reduction phenomenon. In Section~\ref{sec-stab}, we propose a continuous-time feedback controller which almost surely globally exponentially stabilizes the quantum spin-$\frac{1}{2}$ system around the target eigenstate. We also analyze the asymptotic behavior of trajectories associated to the quantum dynamics with feedback. We precise that our continuous feedback has a similar form as the ones used in \cite{van2005feedback, tsumura2008global}. Simulation results provided in Section~\ref{sec-sim} demonstrate the effectiveness of our control design.

\section{Preliminaries}
\label{sec-pre}

\subsection{System description}
Consider a filtered probability space $(\Omega,\mathcal{F},(\mathcal{F}_t),\mathbb{P})$. Let $W_t$ be a standard Wiener process and assume that $\mathcal{F}_t$ is the natural filtration of the process $W_t$. 
The stochastic master equation describing the dynamics of  a quantum spin-$\frac{1}{2}$ system is given by
\begin{equation}
\setlength{\abovedisplayskip}{3pt}
\setlength{\belowdisplayskip}{3pt}
\begin{split}
d\rho_t&=\left(-i\frac{\omega_{eg}}{2}[\sigma_z,\rho_t]+\frac{M}{4}(\sigma_z\rho_t\sigma_z-\rho_t)-i\frac{u_t}{2}[\sigma_y,\rho_t]\right)dt\\
&\quad+\frac{\sqrt{\eta M}}{2}[\sigma_z\rho_t+\rho_t\sigma_z-2\mathrm{Tr}(\sigma_z\rho_t)\rho_t]dW_t\\
&=:F(\rho_t)dt+G(\rho_t)dW_t
\end{split}
\label{2D SME}
\end{equation}
where $u_t = u(\rho_t)$ is the feedback controller, $\omega_{eg}$ is the difference between the energies of the excited state and the ground state, $\eta\in[0,1]$ is determined by the efficiency of the photodetectors, and $M>0$ is the strength of the interaction between the light and the atoms. The matrices $\sigma_x,$ $\sigma_y,$ and $\sigma_z$ correspond to the  Pauli matrices. The quantum state is described by the density operator $\rho$, which belongs to the space $\mathcal{S}$,  
\begin{equation}
\setlength{\abovedisplayskip}{3pt}
\setlength{\belowdisplayskip}{3pt}
\mathcal{S}:=\{\rho\in\mathbb{C}^{2\times 2}:\rho=\rho^*,\mathrm{Tr}(\rho)=1,\rho \geq 0\},
\label{2D State space S}
\end{equation}
where $\rho^*$ denotes Hermitian conjugation. For a 2-level quantum system, $\rho$ can be uniquely characterized by the Bloch sphere coordinates $(x,y,z)$ as 
\begin{equation}
\setlength{\abovedisplayskip}{3pt}
\setlength{\belowdisplayskip}{3pt}
\rho=\frac{\mathds{1}+x\sigma_x+y\sigma_y+z\sigma_z}{2}=\frac12
\begin{bmatrix}
1+z & x-iy\\
x+iy & 1-z
\end{bmatrix}
\label{Density op in Cart}.
\end{equation}
The vector $(x,y,z)$ belongs to the ball
\begin{equation*}
\setlength{\abovedisplayskip}{3pt}
\setlength{\belowdisplayskip}{3pt}
B(\mathbb{R}^3):=\{(x,y,z)\in\mathbb{R}^3: x^2+y^2+z^2 \leq 1 \}.
\end{equation*}
The stochastic differential equation~\eqref{2D SME} expressed in the Bloch sphere coordinates takes the following form
\begin{equation}
\setlength{\abovedisplayskip}{3pt}
\setlength{\belowdisplayskip}{3pt}
\begin{split}
dx_t&=\left(-\omega_{eg} y_t-\frac{M x_t}{2}+u_tz_t \right)dt-\sqrt{\eta M}x_tz_tdW_t\\
dy_t&=\left(\omega_{eg} x_t-\frac{M y_t}{2} \right)dt-\sqrt{\eta M}y_tz_tdW_t \\
dz_t&=-u_tx_tdt+\sqrt{\eta M} \left( 1-z^2_t \right)dW_t.\\
\end{split}
\label{SME Bloch}
\end{equation}
The existence and uniqueness of the solution of~\eqref{2D SME} have been studied in \cite{pellegrini2008existence, mirrahimi2007stabilizing}. 

\subsection{Stochastic stability}
In this subsection, we introduce the notions of stochastic stability  needed throughout the paper by adapting classical notions (see e.g.~\cite{mao2007stochastic, khasminskii2011stochastic}) to our setting. In order to provide them, we first present the definition of Bures distance \cite{bengtsson2017geometry}.
\begin{defi}
The Bures distance  between two quantum states $\rho_a$ and $\rho_b$ lying in the state space $\mathcal{S}$ is given by
\begin{equation*}
\setlength{\abovedisplayskip}{3pt}
\setlength{\belowdisplayskip}{3pt}
d_B(\rho_a,\rho_b) := \sqrt{2-2\sqrt{\mathrm{Tr}(\rho_a\rho_b)+2\sqrt{\det(\rho_a)\det(\rho_b)}}}.
\end{equation*}
Also, the Bures distance between a quantum state $\rho_a$ and a set $E \subseteq \mathcal{S} $ is defined as $d_B(\rho_a, E) = \min_{\rho \in E} d_B(\rho_a,\rho)$.
\end{defi}
Given $E\subset\mathcal{S}$ and $r>0$, we define the neighborhood $B_r(E)$ of $E$ as 
\begin{equation*}
\setlength{\abovedisplayskip}{3pt}
\setlength{\belowdisplayskip}{3pt}
B_r (E) = \{ \rho \in \mathcal{S}: d_B(\rho,E) < r\}.
\end{equation*}
\begin{defi}
Let $\bar E$ be an invariant set of system~\eqref{2D SME}, then $\bar E$ is said to be
\begin{enumerate}
\item[1.] 
\emph{locally stable in probability}, if for every $\epsilon \in (0,1)$ and $r>0$, there exists a $\delta = \delta(\epsilon,r)$, such that,
\begin{equation*}
\setlength{\abovedisplayskip}{3pt}
\setlength{\belowdisplayskip}{3pt}
\mathbb{P} \left\lbrace \rho_t \in B_r (\bar E) \text{ for } t \geqslant 0 \right\rbrace \geq 1-\epsilon,
\end{equation*}
whenever $\rho_0 \in B_{\delta} (\bar E)$.

\item[2.]
\emph{exponentially stable in mean}, if for some positive constants $\alpha$ and $\beta$,
\begin{equation*}
\setlength{\abovedisplayskip}{3pt}
\setlength{\belowdisplayskip}{3pt}
\mathbb{E}\left[  d_B(\rho_t,\bar E) \right] \leq \alpha \,d_B(\rho_0,\bar E)e^{-\beta t},
\end{equation*} 
whenever $\rho_0 \in \mathcal{S}$. 

\item[3.]
\emph{almost surely exponentially stable}, if
\begin{equation*}
\setlength{\abovedisplayskip}{3pt}
\setlength{\belowdisplayskip}{3pt}
\limsup_{t \rightarrow \infty} \frac{1}{t} \log d_B(\rho_t,\bar E) < 0, \quad a.s.
\end{equation*}
whenever $\rho_0 \in \mathcal{S}$. The left-hand side of the above inequality is called the Lyapunov exponent of the solution and its absolute value describes the rate of
convergence.
\end{enumerate}
\end{defi}
Note that any equilibrium $\bar\rho$ of~\eqref{2D SME}, that is any quantum state satisfying $F(\bar\rho)=G(\bar\rho)=0$, corresponds to a special case of invariant set.

\begin{nota}
Suppose that the function $V(\rho,t):\mathcal{S} \times \mathbb{R}_{+} \rightarrow \mathbb{R}_{+}$ is continuously twice differentiable in $\rho$ and once in $t$. We denote by $\mathscr{L}$ the infinitesimal generator  associated with the equation \eqref{2D SME}. The latter acts on $V$ in the following way,
\begin{equation*}
\setlength{\abovedisplayskip}{3pt}
\setlength{\belowdisplayskip}{3pt}
\begin{split}
&\mathscr{L}V(\rho,t) \\
 &:=\frac{\partial V(\rho,t)}{\partial t}+\frac{\partial V(\rho,t)}{\partial \rho}F(\rho)+\frac12 \frac{\partial^2 V(\rho,t)}{\partial \rho^2}(G(\rho),G(\rho)),
 \end{split}
 \end{equation*}
 where derivatives must be taken componentwise and the Hessian must be thought as a quadratic form.
Then, by It\^o's formula, $dV(\rho,t) = \mathscr{L}V(\rho,t)dt+\frac{\partial V(\rho,t)}{\partial \rho}G(\rho)dW_t$.
\end{nota}


\section{Quantum State Reduction}
\label{sec-qsr}
In this section, we study the dynamics of the quantum spin system~\eqref{2D SME} with $u_t=0$. First, we  note that in this case the states 
$\rho_g = \mathrm{diag}(1,0)$ and $\rho_e = \mathrm{diag}(0,1)$ are the equilibria of system~\eqref{2D SME}, i.e., $F(\rho_g)=F(\rho_e)=0$ and $G(\rho_g)=G(\rho_e)=0$. In the following theorem, we  prove that the quantum state reduction of dynamics~\eqref{2D SME} towards the invariant set $\bar E:=\{\rho_g,\rho_e\}$  
  occurs with exponential velocity.
\begin{thm}[Quantum state reduction]
For system~\eqref{2D SME}, with $u_t=0$ and $\rho_0\in \mathcal{S},$ the set $\bar E$ is exponentially stable in mean and a.s. with Lyapunov exponent less than $-\frac{\eta M}{2}$. Moreover, the probability of convergence to $\bar\rho \in \bar E$ is $\mathrm{Tr}(\rho_0 \bar\rho)$.
\label{QSR}
\end{thm}

\proof
The proof proceeds in two steps:

\textbf{Step 1:} 
We take the standard deviation process  
\begin{equation*}
\setlength{\abovedisplayskip}{3pt}
\setlength{\belowdisplayskip}{3pt}
V(\rho_t)=\sqrt{\mathrm{Tr}(\sigma^2_z\rho_t)-\mathrm{Tr}^2(\sigma_z\rho_t)}=\sqrt{1-\mathrm{Tr}^2(\sigma_z\rho_t)}
\end{equation*}
as candidate Lyapunov function.
It is twice continuously differentiable with respect to $\rho_t$, except for the set $\bar E$. By Lemma~\ref{Never reach} given in the appendix, we can assume that $\bar E$ is never attained, which allows us to calculate $\mathscr{L}V(\rho_t)=-\frac{\eta M}{2}V(\rho_t)$.
By It\^o's formula, for all $\rho_0 \in \mathcal{S}$, $\mathbb{E}[V(\rho_t)]=V(\rho_0)+\int^t_0 \mathbb{E}[\mathscr{L}V(\rho_t)]ds$ which implies $\mathbb{E}[V(\rho_t)]=V(\rho_0) e^{-\frac{\eta M}{2}t}$. 
We have
\begin{equation*}
\setlength{\abovedisplayskip}{3pt}
\setlength{\belowdisplayskip}{3pt}
d_B(\rho_t,\bar E)=
\begin{cases}
d_B(\rho_t,\rho_e), &\text{if } \mathrm{Tr}(\sigma_z\rho_t) > 0;\\ 
d_B(\rho_t,\rho_g), &\text{if } \mathrm{Tr}(\sigma_z\rho_t) \leq 0.\\ 
\end{cases}
\end{equation*}
Thus, for all $\rho_t \in \mathcal{S}$, $0 \leq d_B(\rho_t,\bar E) \leq \sqrt{2-\sqrt{2}}$. Moreover,
\begin{equation*}
\setlength{\abovedisplayskip}{3pt}
\setlength{\belowdisplayskip}{3pt}
\begin{split}
&C_1 d_B(\rho_t,\rho_e) \leq V(\rho_t) \leq C_2 d_B(\rho_t,\rho_e),\quad \text{if } \mathrm{Tr}(\sigma_z\rho_t) > 0;\\
&C_1 d_B(\rho_t,\rho_g) \leq V(\rho_t) \leq C_2 d_B(\rho_t,\rho_g),\quad \text{if } \mathrm{Tr}(\sigma_z\rho_t) \leq 0,
\end{split}
\end{equation*}
where $C_1 = \sqrt{1+\sqrt{2}/2}$ and $C_2 = 2$. Then, we have 
\begin{equation}
\setlength{\abovedisplayskip}{3pt}
\setlength{\belowdisplayskip}{3pt}
C_1 d_B(\rho_t,\bar E) \leq V(\rho_t) \leq C_2 d_B(\rho_t,\bar E).
\label{C1d<=V<=C2d}
\end{equation}
Thus, for all $\rho_0 \in \mathcal{S}$, $\mathbb{E}[d(\rho_t,\bar E)] \leq C_2/C_1\,d(\rho_0,\bar E)e^{-\frac{\eta M}{2}t}$, which implies that the set $\bar E$ is exponentially stable in mean.

Now, we consider the following stochastic process,
\begin{equation*}
\setlength{\abovedisplayskip}{3pt}
\setlength{\belowdisplayskip}{3pt}
Q(\rho_t) = e^{\frac{\eta M}{2}t}V(\rho_t) \geq 0,
\end{equation*}
whose infinitesimal generator is given by
\begin{equation*}
\setlength{\abovedisplayskip}{3pt}
\setlength{\belowdisplayskip}{3pt}
\mathscr{L}Q(\rho_t) = e^{\frac{\eta M}{2}t}[ \eta M/2\,V(\rho_t)+\mathscr{L}V(\rho_t) ]=0.
\end{equation*}
Hence, the process $Q(\rho_t)$ is a positive martingale. Due to the Doob's martingale convergence theorem \cite{revuz2013continuous}, the process $Q(\rho_t)$ converges almost surely to a finite limit $A$ as $t \rightarrow \infty$. Consequently, $\sup_{t \geq 0}Q(\rho_t) = A$ implies $\sup_{t \geq 0} V(\rho_t) = Ae^{-\frac{\eta M}{2}t}$ a.s.. Letting $t \rightarrow \infty$, we obtain: $\limsup_{t \rightarrow \infty} \frac{1}{t} \log V(\rho_t) \leq -\frac{\eta M}{2}$ a.s.. By the inequality~\eqref{C1d<=V<=C2d},
\begin{equation} 
\setlength{\abovedisplayskip}{3pt}
\setlength{\belowdisplayskip}{3pt}
\limsup_{t\rightarrow\infty}\frac{1}{t}\log d_B(\rho_t,\bar E) \leq -\frac{\eta M}{2}, \qquad a.s.
\label{rate QSR}
\end{equation}
which means that the set $\bar E$ is a.s. exponentially stable. Furthermore, we conclude that almost all paths which never exit the set $B_r(\bar\rho) \subset \mathcal{S}$ exponentially converge to $\bar\rho \in \bar E$.

\textbf{Step 2:}
We follow the approach of \cite{amini2013feedback, adler2001martingale} to calculate the probability of convergence towards $\bar\rho \in \bar E$. According to the result of Step 1, we denote the terminal state $\rho_{\infty}$ of the reduction process $\rho_{\infty} := P_e\rho_e+P_g\rho_g$, where $P_e$ (resp. $P_g$) is the probability of reducing to $\rho_e$ (resp. $\rho_g$), as $t \rightarrow \infty$. Since $\mathscr{L}\mathrm{Tr}(\rho_t \bar\rho)=0$, $\mathrm{Tr}(\rho_t \bar\rho)$ is a positive martingale. By the Doob's martingale convergence theorem, we have
\begin{equation*}
\setlength{\abovedisplayskip}{3pt}
\setlength{\belowdisplayskip}{3pt}
\mathbb{E}[\mathrm{Tr}(\rho_{\infty} \bar\rho)]=\lim_{t\rightarrow\infty}\mathbb{E}[\mathrm{Tr}(\rho_t\bar\rho)]=\mathbb{E}[\mathrm{Tr}(\rho_0 \bar\rho)]=\mathrm{Tr}(\rho_0 \bar\rho).
\end{equation*}
Therefore, $P_e=\mathrm{Tr}(\rho_0\rho_e)$, $P_g=\mathrm{Tr}(\rho_0\rho_g)$. The proof is complete.\hfill$\square$

\section{Exponential Stabilization by Continuous Feedback}
\label{sec-stab}

\subsection{Almost sure global exponential stabilization}
\label{sec-stab-1}

\begin{lem}
For system~\eqref{2D SME} with $\rho_0 \in \mathcal{S}$, assume that under the feedback law $u_t$ the system~\eqref{2D SME} admits $\bar\rho \in \bar E$ as unique equilibrium, then for all $r \in (0,\sqrt{2}]$, 
\begin{equation*}
\setlength{\abovedisplayskip}{3pt}
\setlength{\belowdisplayskip}{3pt}
\mathbb{P}_{\rho_0}\{\tau_{B_r} < \infty\}=1,
\end{equation*}
where $\tau_{B_r}$ denotes the first entry time of $B_r(\bar{\rho})$.
\label{Passage lemma}
\end{lem}

\proof
In order to prove the lemma, we will make use of the support theorem, in the form considered in~\cite{baxendale1991invariant}. For this purpose, one has to consider the Stratonovich version of the equation~\eqref{2D SME} and the corresponding controlled deterministic differential equation, which takes the form 
\begin{equation}
\label{eq-deterministic}
\setlength{\abovedisplayskip}{3pt}
\setlength{\belowdisplayskip}{3pt}
\dot \rho=F(\rho)-\frac12 \frac{\partial G(\rho)}{\partial \rho} G(\rho)+G(\rho)w(t),
\end{equation} 
where $w$ is the (deterministic) control input. Assume without loss of generality that $\bar \rho = \rho_e$, then by the hypothesis of the lemma we have that $F(\rho_g)\neq 0$ and therefore $\rho_g$ is not an equilibrium point of~\eqref{eq-deterministic}. Moreover, since the $z$ component of $G(\rho)$ in the Bloch sphere coordinates is different from zero outside $\bar E$, we deduce that, for any $\rho(0)\in  \mathcal{S}$, there exists a control $w(t)$ steering~\eqref{eq-deterministic} from $\rho(0)$ to $B_r(\bar{\rho})$ in finite time. Thus, the assumptions in~\cite[Propositions 4.3 and 4.6]{baxendale1991invariant} are satisfied, in  such a way that, for a fixed $T>0$, there exists $\epsilon_T>0$ such that, for all $\rho_0 \in \mathcal{S}$, $\mathbb{P}_{\rho_0}(\tau_{B_r} < T) \geq \epsilon_T$. 


By Dynkin's inequality \cite{dynkin1965markov}, $\forall \, \rho_0 \in \mathcal{S}$ and $\forall \, r\in(0,\sqrt{2}]$,
\begin{equation*}
\setlength{\abovedisplayskip}{3pt}
\setlength{\belowdisplayskip}{3pt}
\mathbb{E}_{\rho_0}[\tau_{B_r}] \leq \frac{T}{1-\sup_{\rho_0 \in \mathcal{S}}\mathbb{P}_{\rho_0}(\tau_{B_r} \geq T)} \leq \frac{T}{\epsilon_{T}}<\infty.
\end{equation*}
Then by Markov inequality, $\forall \, \rho_0 \in \mathcal{S}$ and $\forall \, r\in(0,\sqrt{2}]$, $\mathbb{P}_{\rho_0}\{\tau_{B_r}<\infty\}=1$. 
The proof is complete.\hfill$\square$
 
\begin{thm}
For system~\eqref{2D SME} with $\rho_0 \in \mathcal{S}$, denote as $\bar\rho \in \bar E$ the target eigenstate, and assume that the feedback controller $u_t$ satisfies the condition of Lemma~\ref{Never reach} given in the appendix and $u_t = 0$ in $ \bar E$ if and only if $\rho_t = \bar\rho$. Suppose that there exists a function $V(\rho)$, which is twice continuously differentiable on the set $B_r(\bar\rho)\setminus\bar\rho$ with $r \in (0,\sqrt{2}]$, and positive constants $C$, $C_1$, $C_2$ such that, $\forall \, \rho_t \in B_r(\bar\rho)\setminus\bar\rho$
\begin{enumerate}[\topsep=0pt,\itemsep=0pt]
\item[(i)] $C_1 \, d_B(\rho_t,\bar\rho) \leq V(\rho_t) \leq C_2 \, d_B(\rho_t,\bar\rho)$,
\item[(ii)] $\mathscr{L} V(\rho_t) \leq -C \, V(\rho_t)$.
\end{enumerate}
Then $\bar\rho$ is a.s. exponentially stable.
\label{Th feedback}
\end{thm}

\proof
The proof proceeds in 3 steps:
\begin{enumerate}
\item[1.] $\bar\rho$ is locally stable in probability.

\item[2.] $\exists \, T<\infty$ s.t. $\forall \, t \geq T$, $\rho_t \in B_r(\bar\rho)$ a.s..

\item[3.] $\bar\rho$ is a.s. exponentially stable.
\end{enumerate}

\textbf{Step 1.} 
Choose $r\in (0,\sqrt{2})$ and let $\epsilon \in (0,1)$ be arbitrary, by the continuity of $V(\rho_t)$ and the fact that, for all $\rho_t \in \mathcal{S}$, $V(\rho_t)=0$ if and only if $d_B(\rho_t,\bar\rho)=0$, we can find a $\delta=\delta(\epsilon,r)>0$ such that,
\begin{equation}
\setlength{\abovedisplayskip}{3pt}
\setlength{\belowdisplayskip}{3pt}
1/\epsilon \sup_{\rho_0 \in B_{\delta}(\bar\rho)}V(\rho_0) \leq C_1r.
\label{Step2:stable ineq}
\end{equation} 
Let $\tau$ be the first exit time of $\rho_t$ from $B_r(\bar\rho)$. By It\^o's formula and the condition (ii),
\begin{equation*}
\setlength{\abovedisplayskip}{3pt}
\setlength{\belowdisplayskip}{3pt}
\mathbb{E}[V(\rho_{t \wedge \tau})] \leq V(\rho_0)-C \, \mathbb{E} \left[\int^{t \wedge \tau}_0 V(\rho_s)ds \right] \leq V(\rho_0).
\end{equation*}
For all $t \geq \tau$, $d_B(\rho_{t \wedge \tau},\bar\rho)=d_B(\rho_{\tau},\bar\rho)=r$. Hence, by the condition (i),
\begin{equation*}
\setlength{\abovedisplayskip}{3pt}
\setlength{\belowdisplayskip}{3pt}
\begin{split}
&\mathbb{E}[V(\rho_{t \wedge \tau})] \geq \mathbb{E}[\mathds{1}_{\{t \wedge \tau\}}V(\rho_{\tau})] \\
&\qquad \geq \mathbb{E}[\mathds{1}_{\{t \wedge \tau\}}C_1d_B(\rho_{\tau},\bar\rho)] \geq C_1 r \, \mathbb{P}(\tau \leq t).
\end{split}
\end{equation*}
Combining with the inequality~\eqref{Step2:stable ineq}, we have 
\begin{equation*}
\setlength{\abovedisplayskip}{3pt}
\setlength{\belowdisplayskip}{3pt}
\mathbb{P}(\tau \leq t) \leq \frac{\mathbb{E}[V(\rho_{t \wedge \tau})]}{C_1r} \leq \frac{V(\rho_{0})}{C_1r} \leq \epsilon.
\end{equation*}
Letting $t \rightarrow \infty$, 
\begin{equation*}
\setlength{\abovedisplayskip}{3pt}
\setlength{\belowdisplayskip}{3pt}
\mathbb{P} (\tau<\infty) \leq \epsilon \Rightarrow \mathbb{P} \{ d_B(\rho_t,\bar\rho) < r \text{ for } t \geq 0 \} \geq 1-\epsilon,
\end{equation*}
which concludes the first step.

\textbf{Step 2.}
Since $u_t=0$ in $\bar E$ if and only if $\rho_t=\bar\rho$, by Lemma~\ref{Passage lemma}, we obtain, for all $\rho_0 \in \mathcal{S}$, $\mathbb{P}_{\rho_0}\{\tau_{B_{\delta}} < \infty\}=1$, where $\tau_{B_{\delta}} = \inf\{t \geq 0: \rho_t \in B_{\delta}(\bar{\rho}) \}$ with $\delta \in(0,r)$. It implies that $\rho_t$ enters $B_{\delta}(\bar{\rho})$ in a finite time almost surely. Then we employ an argument inspired by \cite{mirrahimi2007stabilizing} and suppose $\rho_0 \in \mathcal{S}$. We define two sequences of stopping times $\{\sigma_k\}_{k\geq 0}$ and $\{\tau_k\}_{k\geq 1}$ with $\sigma_0=0$ and, for $n \geq 1$, $\tau_n = \inf\{t \geq \sigma_{n-1}: \rho_t \in B_{\delta}(\bar{\rho})\}$ and $\sigma_n = \inf\{t \geq \tau_{n}:\rho_t \notin B_r(\bar{\rho})\}$. Then by the strong Markov property and the Bayes' formula, we have
\begin{equation*}
\setlength{\abovedisplayskip}{3pt}
\setlength{\belowdisplayskip}{3pt}
\mathbb{P}_{\rho_0}(\tau_n < \infty|\sigma_{n-1} < \infty)=1,\; \mathbb{P}_{\rho_0}(\sigma_n < \infty|\tau_n < \infty)\leq \epsilon.
\end{equation*}
Moreover, by the construction of the stopping times, we have
\begin{equation*}
\setlength{\abovedisplayskip}{3pt}
\setlength{\belowdisplayskip}{3pt}
\mathbb{P}_{\rho_0}(\tau_n < \infty|\sigma_n < \infty)=\mathbb{P}_{\rho_0}(\sigma_{n-1} < \infty|\tau_n < \infty)=1.
\end{equation*}
Hence by Bayes' formula, we obtain
\begin{align*}
&\frac{\mathbb{P}_{\rho_0}(\sigma_n < \infty)}{\mathbb{P}_{\rho_0}(\sigma_{n-1} < \infty)}\\
& =\mathbb{P}_{\rho_0}(\sigma_n < \infty|\tau_n < \infty)
\mathbb{P}_{\rho_0}(\tau_n < \infty|\sigma_{n-1} < \infty).
\end{align*}
In addition with $\mathbb{P}_{\rho_0}(\sigma_1 < \infty) \leq \epsilon$, we have $\mathbb{P}_{\rho_0}(\sigma_n < \infty) \leq \epsilon^n$ and thus $\sum^{\infty}_{n=1}\mathbb{P}_{\rho_0}(\sigma_n < \infty) \leq \sum^{\infty}_{n=1}\epsilon^n = \frac{\epsilon}{1-\epsilon}<\infty$. By the Borel-Cantelli lemma, we can conclude that, for almost all trajectories, there exists $N<\infty$, such that $\sigma_n=\infty$ for all $n \geq N$. That is, for almost all sample paths, there exists $T<\infty$ such that, for all $t \geq T$, $\rho_t \in B_r(\bar{\rho})$. This completes the proof of Step~2. 

\textbf{Step 3.}
In this step, we obtain an upper bound of the Lyapunov exponent by employing an argument inspired by~\cite[Theorem 3.3, p.~121]{mao2007stochastic}.
For every fixed $T$ consider the event $\Omega_T=\{\rho_t \in  B_r(\bar\rho) \text{ for all } t\geq T\}$. Conditioning to $\rho_t\in \Omega_T$, from the condition (ii), we get
\begin{equation*}
\setlength{\abovedisplayskip}{3pt}
\setlength{\belowdisplayskip}{3pt}
\mathscr{L}\log V(\rho_t) \leq -C-\frac{1}{2} [V_\rho(\rho_t) G(\rho_t)/V(\rho_t) ]^2 =: -C-\frac{g^2(t)}{2}
\end{equation*} 
for $t\geq T$ which implies
\begin{equation*}
\setlength{\abovedisplayskip}{3pt}
\setlength{\belowdisplayskip}{3pt}
\begin{split}
\log V(\rho_t)=\log V(\rho_{T})- &C(t-T)\\
&+\int^t_{T}g(s)dW_s -\frac12\int^t_{T}g^2(s)ds.
\end{split}
\end{equation*}

Let $m=1,2,3\cdots$ and take arbitrarily $\epsilon\in(0,1)$. By the exponential martingale inequality (see e.g.~\cite{mao2007stochastic}), we have
\begin{equation*}
\setlength{\abovedisplayskip}{3pt}
\setlength{\belowdisplayskip}{3pt}
\mathbb{P} \left\lbrace\! \sup_{T \leq t\leq T+m} \!\left[ \int^t_{T}\!\!g(s)dW_s-\frac{\epsilon}{2}\!\int^t_{T}\!\!g^2(s)ds \right] \!> \!\frac{2}{\epsilon} \log m \right\rbrace \!\leq\! \frac{1}{m^2}
\end{equation*}
Since $\sum^{\infty}_{m=1}\frac{1}{m^2}<\infty$, by Borel-Cantelli lemma, we have that for almost all sample paths, there exists $m_0$, such that if $m>m_0$,
\begin{equation*}
\setlength{\abovedisplayskip}{3pt}
\setlength{\belowdisplayskip}{3pt}
\sup_{T\leq t\leq T+m}\left[ \int^t_{T}g(s)dW_s-\frac{\epsilon}{2}\int^t_{T}g^2(s)ds \right] \leq\frac{2}{\epsilon}\log m
\end{equation*}
Thus, for $T\leq t\leq T+m$ and $m>m_0$,
\begin{equation*}
\setlength{\abovedisplayskip}{3pt}
\setlength{\belowdisplayskip}{3pt}
\int^t_{T}g(s)dW_s \leq \frac{2}{\epsilon}\log m+\frac{\epsilon}{2}\int^t_{T}g^2(s)ds,\quad a.s.
\end{equation*}
If $g^2(t) \geq K$, then we have
\begin{equation*}
\setlength{\abovedisplayskip}{3pt}
\setlength{\belowdisplayskip}{3pt}
\log V(\rho_t) \leq \log V(\rho_{T})-C(t-T)+\frac{2}{\epsilon}\log m-\frac{1-\epsilon}{2}K
\end{equation*}
almost surely. It gives
\begin{equation*} 
\setlength{\abovedisplayskip}{3pt}
\setlength{\belowdisplayskip}{3pt}
\limsup_{t \rightarrow \infty}\frac{1}{t}\log V(\rho_t) \leq -C-\frac{1-\epsilon}{2}K \quad a.s.
\end{equation*}
Letting $\epsilon \rightarrow 0$, we have $\limsup_{t \rightarrow \infty}\frac{1}{t}\log V(\rho_t) \leq -C-\frac{K}{2}$ a.s.
In addition, due to (i), we have
\begin{equation*}
\setlength{\abovedisplayskip}{3pt}
\setlength{\belowdisplayskip}{3pt}
\limsup_{t\rightarrow\infty}\frac{1}{t}\log d_B(\rho_t,\bar\rho) \leq -C-\frac{K}{2},\quad a.s.
\end{equation*}
Since $T$ can be taken arbitrarily large and Step 2 implies that $\lim_{T\to \infty}\mathbb{P}(\Omega_{T})=1$ we can conclude that $\bar\rho$ is almost surely exponentially stable. The proof is complete.\hfill$\square$

\subsection{Feedback controller design}
\label{sec-stab-2}

\subsubsection{Control design}
 \label{sec-stab-2A}

For simplicity, we define the following subset of $\mathcal{S}$,
\begin{equation*}
\setlength{\abovedisplayskip}{3pt}
\setlength{\belowdisplayskip}{3pt}
D_{\lambda}(\bar\rho) := \{\rho \in \mathcal{S}:0 < \lambda < \mathrm{Tr}(\rho\bar\rho) \leq 1\} = B_{r_{\lambda}}(\bar\rho)
\end{equation*}
where $r_{\lambda} = \sqrt{2-2\sqrt{\lambda}}$ with $\lambda \in (0,1)$. 

\begin{thm}
Consider system~\eqref{2D SME} with $\rho_{0} \in \mathcal{S}$,  denote as $\bar\rho \in \bar E$ the target eigenstate and let 
\begin{equation}
\setlength{\abovedisplayskip}{3pt}
\setlength{\belowdisplayskip}{3pt}
V(\rho_t) = \sqrt{1-\mathrm{Tr}(\rho_t\bar\rho)},\quad \bar\rho \in \bar E.
\label{Lya fun}
\end{equation}
Define the feedback control
\begin{equation}
\setlength{\abovedisplayskip}{3pt}
\setlength{\belowdisplayskip}{3pt}
u_t = u^{(1)}_t + u^{(2)}_t = \alpha [V(\rho_t)]^{\beta} - \gamma \, \mathrm{Tr}(i[\sigma_y,\rho_t]\bar\rho)
\label{u_t}
\end{equation}
where $\gamma \geq 0$, $\beta \geq 1$ and $0< \alpha <\frac{\eta M\lambda^2}{(1-\lambda)^{\frac{\beta-1}{2}}}$ with $\lambda \in (0,1)$.
Then $u_t$ globally exponentially stabilizes system~\eqref{2D SME} a.s. to $\bar\rho$.
\end{thm}

\proof
As  $\rho_t \geq 0,$ we have 
\begin{equation*}
\setlength{\abovedisplayskip}{3pt}
\setlength{\belowdisplayskip}{3pt}
|\mathrm{Tr}(i[\sigma_y,\rho_t]\bar{\rho})|=|\mathrm{Tr}(\sigma_x\rho_t)|=|x_t| \leq 2V(\rho_t).
\end{equation*}
We easily verify that $u_t$ satisfies the condition of Theorem~\ref{Th feedback}. The infinitesimal generator of~\eqref{Lya fun} is given by
\begin{equation}
\setlength{\abovedisplayskip}{3pt}
\setlength{\belowdisplayskip}{3pt}
\mathscr{L}V(\rho_t)=\frac{u_t}{4}\frac{\mathrm{Tr}(i[\sigma_y,\rho_t]\bar\rho)}{V(\rho_t)}-\frac{\eta M}{2}\mathrm{Tr}^2(\rho_t\bar\rho)V(\rho_t)
\label{infi gen Lya fun}
\end{equation}
For all $\rho_t \in D_{\lambda}(\bar\rho)$, 
\begin{equation*}
\setlength{\abovedisplayskip}{3pt}
\setlength{\belowdisplayskip}{3pt}
\begin{split}
\mathscr{L}V(\rho_t) &\leq -1/2[ \eta M \lambda^2-\alpha V^{\beta-1}(\rho_t) ] V(\rho_t)\\
&\leq -1/2[ \eta M \lambda^2-\alpha(1-\lambda)^{\frac{\beta-1}{2}} ] V(\rho_t)\\
\end{split}
\end{equation*}
If $0 < \alpha < \frac{\eta M\lambda^2}{(1-\lambda)^{\frac{\beta-1}{2}}}$, we get
\begin{equation}
\setlength{\abovedisplayskip}{3pt}
\setlength{\belowdisplayskip}{3pt}
\mathscr{L}V(\rho_t) \leq -C_{\lambda} V(\rho_t)
\label{LVn<=-CVn}
\end{equation}
where $C_{\lambda} = 1/2[\eta M \lambda^2-\alpha(1-\lambda)^{\frac{\beta-1}{2}}] > 0$. Moreover, we have
\begin{equation*}
\setlength{\abovedisplayskip}{3pt}
\setlength{\belowdisplayskip}{3pt}
g^2(t) = [V_\rho(\rho_t) G(\rho_t)/V(\rho_t) ]^2 \geq \eta M \lambda^2, \quad \forall \, \rho_t \in D_{\lambda}(\bar\rho)
\end{equation*}
By Theorem~\ref{Th feedback}, we can show that $u_t$ exponentially stabilizes the system~\eqref{2D SME} to $\bar \rho$ 
almost surely. Furthermore, we have
\begin{equation*}
\setlength{\abovedisplayskip}{3pt}
\setlength{\belowdisplayskip}{3pt}
\limsup_{t\rightarrow\infty}\frac{1}{t}\log d_B(\rho_t,\rho_n) \leq -K_{\lambda},\quad a.s.
\end{equation*}
where $K_{\lambda} = \eta M \lambda^2-\frac{\alpha}{2}(1-\lambda)^{\frac{\beta-1}{2}}$. The proof is complete.
\phantom{a}\hfill$\square$

\subsubsection{Effect of the parameters of the controller}
 \label{sec-stab-2B}

We now study the dynamics of the closed-loop system and discuss informally how to choose the coefficients $\alpha,\beta,\gamma$ of $u_t$ to stabilize the system more efficiently.

\textbf{Case~1.} Consider the case in which the trajectory $\rho_t$ lies in a small neighborhood of the target eigenstate $\bar\rho$ permanently. According to the infinitesimal generator of the Lyapunov function~\eqref{Lya fun} given by~\eqref{infi gen Lya fun}, we can see that the stochastic part of the stochastic master equation~\eqref{2D SME} and the deterministic part multiplied by $u^{(2)}_t$ always yield desirable effects on the preparation of the target eigenstate. On the other hand $u^{(1)}_t$ may impact negatively on the convergence to the target state. Choosing a larger $\beta$ allows us  to reduce the negative effect of $u^{(1)}_t$ for any fixed $\alpha$.

\textbf{Case~2.} Assume that the initial state $\rho_{0}$ is contained in a neighborhood of the antipodal eigenstate. 
It is easy to check that, if such a neighborhood is small enough, then $|\mathscr{L}x_t|\geq \frac{\alpha}2$. In particular, choosing a larger $\alpha$ allows one to consider a larger neighborhood such that the inequality holds true. Letting $\tau$ be the first exit time from the neighborhood, by It\^o's formula, we have $2 \geq \mathbb{E}[|x_{\tau} - x_0|] \geq \frac{\alpha}2 \mathbb{E}[\tau]$. By Markov's inequality, we then get $\mathbb{P}[\tau\geq t]\leq \frac {\mathbb{E}[\tau]}{t}\leq \frac4{\alpha t}$. Thus taking a larger $\alpha$ permits to exit more quickly from the given neighborhood.

\textbf{Case~3.} Suppose that $\rho_0$ is far from $\bar E$.
 In this case, taking $V$ as in~\eqref{Lya fun}, the term in $\mathscr{L}V$ corresponding to $u^{(2)}_t$ is always negative outside a neighborhood $U$ of the antipodal eigenstate, and it is proportional to $\gamma$. Thus, reasoning as in Case~2 above, restricting to the trajectories that never enter $U$ for $t \geq 0$ and defining $\tau$ as the first time when the trajectory enters in a neighborhood $U'$ of $\bar \rho$, we get that a larger $\gamma$ increases the probability of quickly entering $U'$.

\section{Numerical Simulations}
\label{sec-sim}
For our simulations, since $\rho_t$ should remain in $\mathcal{S},$ we consider the following Kraus form \cite{amini2011stability, rouchon2014models} associated with~\eqref{2D SME}, 
\begin{equation*}
\setlength{\abovedisplayskip}{3pt}
\setlength{\belowdisplayskip}{3pt}
\rho_t+d\rho_t=\frac{\mathbb{M}_{dY_t}\rho_t\mathbb{M}^*_{dY_t}+\frac{1-\eta M}{4}\sigma_z\rho_t\sigma_zdt}{\mathrm{Tr}\left( \mathbb{M}_{dY_t}\rho_t\mathbb{M}^*_{dY_t}+\frac{1-\eta M}{4}\sigma_z\rho_t\sigma_zdt \right)},
\end{equation*}
where 
\begin{equation*}
\setlength{\abovedisplayskip}{3pt}
\setlength{\belowdisplayskip}{3pt}
\begin{split}
\mathbb{M}_{dY_t}&=\mathds{1}-\left[ \frac{i}{2}(\omega_{eg}\sigma_z+u_t\sigma_y)+\frac{M}{8}\mathds{1} \right]dt+\frac{\sqrt{\eta M}}{2}\sigma_zdY_t\\
dY_t&=dW_t+\sqrt{\eta M}\mathrm{Tr}(\sigma_z\rho_t)dt
\end{split}
\end{equation*}
\begin{figure}[thpb]
      \includegraphics[width=9cm]{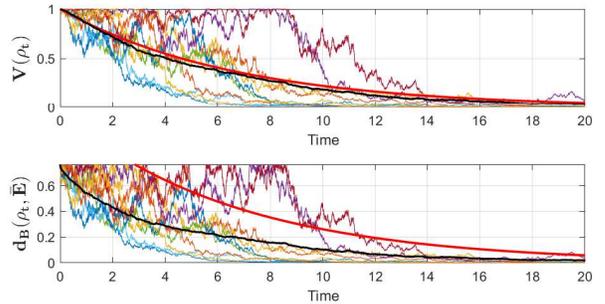}
      \caption{Quantum state reduction with $u_t=0$ starting at $(0,0,0)$, when $\omega_{eg}=0$, $\eta=0,3$, $M=1$: The black curve represents the mean value of the 10 arbitrary samples, the red curve represents the exponential reference.}
      \label{QSRfig}
   \end{figure}
The simulations in the case $u_t=0$ are shown in  Fig.~\ref{QSRfig}. 
In particular, we observe that the expectation of the Lyapunov function $\mathbb{E}[V(\rho_t)]$ fits precisely the exponential function $V(\rho_0)e^{-\frac{\eta M}{2}t}$, and the expectation of the Bures distance $\mathbb{E}[d_B(\rho_t,\bar{E})]$ is always below the exponential function $\frac{2d_B(\rho_0,\bar{E})}{\sqrt{1+\sqrt{2}/2}}e^{-\frac{\eta M}{2}t}$, which confirms the results of Section~\ref{sec-qsr}.
   
\begin{figure}[thpb]
      \includegraphics[width=9.5cm]{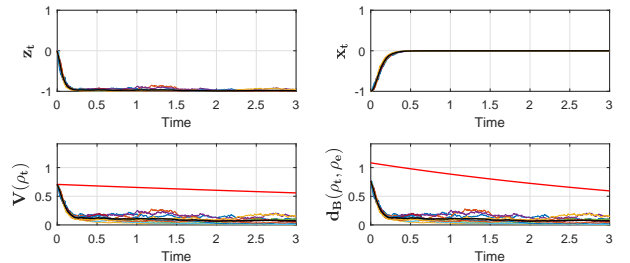}
      \caption{Exponential stabilization with $u_t$ starting at $(-1,0,0)$, when $\omega_{eg}=0$, $\eta=0,3$, $M=1$, $\alpha=7.61$, $\beta=5$, $\lambda=0.9$, $\gamma=10$: The black curve represents the mean value of all the samples, the red curve represents the exponential reference.}
      \label{FB1fig}
\end{figure}

\begin{figure}[thpb]
      \vspace{-0.2cm}
      \setlength{\abovecaptionskip}{-0.2cm}
      \setlength{\belowcaptionskip}{-1.2cm}
      \includegraphics[width=9.5cm]{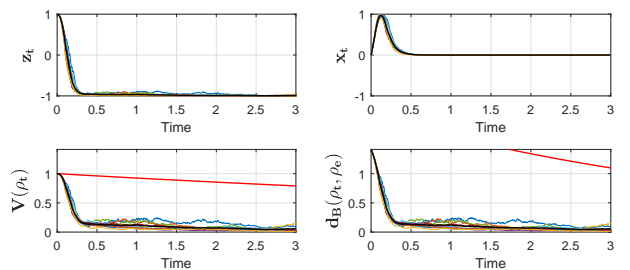}
      \caption{Exponential stabilization with $u_t$ starting at $(0,0,1)$, when $\omega_{eg}=0$, $\eta=0,3$, $M=1$, $\alpha=7.61$, $\beta=5$, $\lambda=0.9$, $\gamma=10$: The black curve represents the mean value of all the samples, the red curve represents the exponential reference.}
      \label{FB2fig}
\end{figure}

We set $\rho_e$ as the target eigenstate. The simulations corresponding to the exponential stabilization with $u_t$ considered as in Section~\ref{sec-stab-2}, are shown in Fig.~\ref{FB1fig} and~\ref{FB2fig}.
In Fig.~\ref{FB1fig}, the system starting at $(-1,0,0)$ satisfies the condition of Case~3 of Section~\ref{sec-stab-2B}. We observe that $V(\rho_t)$ and $d_B(\rho_t,\rho_e)$ decrease quickly, that is $\rho_t$ approaches $\rho_e$ quickly. In Fig.~\ref{FB2fig}, the system starting at $(0,0,1)$ satisfies the condition of Case~2 and Case~3 in Section~\ref{sec-stab-2B}. We observe that $V(\rho_t)$ and $d_B(\rho_t,\rho_e)$ also decrease quickly. 
\section{CONCLUSIONS AND FUTURE WORKS}
\label{sec-conc}
In this paper, we have analyzed the asymptotic behavior of quantum spin-$\frac{1}{2}$ systems both for the cases without and with feedback. For the case without feedback, we have shown that for almost all trajectories $\rho_t$ exponentially converge to the eigenstate $\rho_e$ or $\rho_g$ in mean and almost surely. By combining such asymptotic behavior for the case without feedback and the idea of breaking the symmetry of the state space $\mathcal{S}$, we designed a parametrized continuous feedback stabilizing exponentially the system toward a predetermined target eigenstate almost surely. Furthermore, we established lower bounds both with and without feedback. Also, we discussed the role that each parameter appearing in the feedback law plays in this convergence speed. Finally, numerical simulations confirm our theoretical results. 

Future work will be focused on the extension of the results presented in this paper to the exponential stabilization of an arbitrary eigenstate of $N$-level angular momentum system. Different directions remain to be explored. For example, optimizing our choice of feedback. Also, the adaptation of our results to the case where there are delays in the feedback loop is included in our future research lines.   
\section{ACKNOWLEDGMENTS}
The authors would like to thank M. Mirrahimi, P. Rouchon, and A. Sarlette for stimulating discussions. This work was supported by the Agence Nationale de la Recherche project QUACO ANR-17-CE40-0007.

\appendix

\section{Appendix}
The following lemma is inspired by analogous results in~\cite{khasminskii2011stochastic, mao2007stochastic}. 

\begin{lem}
Assume that $\rho_0 \neq \bar\rho$ with $\bar\rho\in \bar E$ and $u_t$ is continuous, continuously differentiable in $\mathcal S\setminus \bar \rho$ and $|u_t| \leq \Gamma \sqrt{1-\mathrm{Tr}(\rho_t\bar\rho)}$ for some  $\Gamma \in \mathbb{R}_+$. Then,
$\mathbb{P}\{ \rho_t \neq \bar\rho, \forall t\geq 0 \}=1.$
\label{Never reach}
\end{lem}

\proof
Due to the fact that $\rho_t \geq 0$, in the Bloch sphere coordinates, if $|z_t| \rightarrow 1$ then $(x_t,y_t) \rightarrow (0,0)$. Thus, it is sufficient to show the inaccessibility of $\rho_e$ or $\rho_g$ by proving the following two assertions
\begin{enumerate}
\item
For all $z_0 \neq 1$, $|u_t| \leq \Gamma\sqrt{1-z_t}$ with $\Gamma \in \mathbb{R}_+$,
\begin{equation*}
\setlength{\abovedisplayskip}{3pt}
\setlength{\belowdisplayskip}{3pt}
\mathbb{P}\{z_t \neq 1, \forall t\geq 0\}=1.
\end{equation*}
\item
For all $z_0 \neq -1$, $|u_t| \leq \Gamma\sqrt{1+z_t}$ with $\Gamma \in \mathbb{R}_+$,
\begin{equation*}
\setlength{\abovedisplayskip}{3pt}
\setlength{\belowdisplayskip}{3pt}
\mathbb{P}\{z_t \neq -1, \forall t\geq 0\}=1.
\end{equation*}
\end{enumerate}

We will prove the first assertion. 
Suppose $\{z_t=1\}$ is accessible, that is there exists $z_0$ such that $\mathbb{P}\{\tau<\infty\}>0$, where $\tau=\inf\{t \geq 0:z_t=1\}$. Then, we can find a finite constant $T$ sufficiently large such that $\mathbb{P}(B)>0$, where $B=\{\tau \leq T\}$. Consider the stochastic process, $Q_t=1/(1-z_t)$ for all $z_t \neq 1$ and $t\in[0,T]$. By It\^o's formula, 
\begin{equation*}
\setlength{\abovedisplayskip}{3pt}
\setlength{\belowdisplayskip}{3pt}
\mathscr{L}Q_t=-\frac{u_tx_t}{(1-z_t)^2}+\frac{\eta M \left( 1-z^2_t \right)^2}{(1-z_t)^2}\leq KQ_t
\end{equation*}
where $K=\sqrt{2} \, \Gamma+4\eta M$. The above inequality holds as 
\begin{equation*}
\setlength{\abovedisplayskip}{3pt}
\setlength{\belowdisplayskip}{3pt}
\left|\sqrt{\eta M}(1-z_t^2)\right| \leq 2\sqrt{\eta M}(1-z_t) 
\end{equation*} and, by the assumption of the lemma, $|u_tx_t|\leq \Gamma\sqrt 2(1-z_t).$

Due to the Gr\"onwall-Bellman lemma, we construct the following process, $f_t=e^{-Kt}Q_t$ whose infinitesimal generator is given by $$\setlength{\abovedisplayskip}{3pt}
\setlength{\belowdisplayskip}{3pt}
\mathscr{L}f_t=e^{-Kt}(\mathscr{L}Q_t-KQ_t).$$ Now for any $\epsilon \in (z_0,1)$, define the stopping time $\tau_{\epsilon}=\inf\{t \geq 0:z_t \notin [-1,\epsilon)\}$. By It\^o's formula, 
\begin{equation*}
\setlength{\abovedisplayskip}{3pt}
\setlength{\belowdisplayskip}{3pt}
\mathbb{E}[f_{\tau_{\epsilon} \wedge t}]=Q_0+\mathbb{E}\left[\int^{\tau_{\epsilon} \wedge t}_0\mathscr{L}f_sds\right]\leq \frac{1}{1-z_0}.
\end{equation*}
For all events in $B$, we have $\tau_{\epsilon} \leq T$ and $z_{\tau_{\epsilon}}=\epsilon$,
\begin{equation*}
\setlength{\abovedisplayskip}{3pt}
\setlength{\belowdisplayskip}{3pt}
\mathbb{E}\left[\frac{e^{-KT}}{1-\epsilon}\mathds{1}_B\right] \leq \mathbb{E}[f_{\tau_\epsilon}\mathds{1}_B] \leq \frac{1}{1-z_0}
\end{equation*}
which implies $\mathbb{P}(B) \leq (1-\epsilon)e^{-KT}/(1-z_0)$. Letting $\epsilon \rightarrow 1,$ we get $\mathbb{P}(B)=0,$ which contradicts the definition of $B$. Hence, the first assertion is proved.

For the second assertion, we have $|u_tx_t| \leq \Gamma\sqrt{2}(1+z_t)$ and $\left|\sqrt{\eta M}(1-z^2_t)\right| \leq 2\sqrt{\eta M}(1+z_t)$. We can show similarly to the previous case that $\{z_t=-1\}$ is inaccessible. The proof is then complete.
\phantom{a}\hfill$\square$




\bibliography{LIANG_CDC}
\bibliographystyle{plain}
\end{document}